\documentclass[a4paper,10pt]{iopart}

\usepackage{amssymb}
\usepackage{graphicx}
\usepackage{xcolor}
\usepackage[utf8]{inputenc}
\usepackage{hyperref}

\newcommand{\ex}[1]    {\ensuremath{\times 10^{#1}}}

\definecolor{Sr1color}{RGB}{64,142,211}
\definecolor{Sr2color}{RGB}{246,135,18}
\definecolor{SrDcolor}{RGB}{156,22,97}
\definecolor{colorSYRTE} {RGB}{  0,128,  0}
\definecolor{colorJILA}  {RGB}{255,  0,  0}
\definecolor{colorTokyo} {RGB}{ 18,152,188}
\definecolor{colorPTB}   {RGB}{255,128,  0}
\definecolor{colorNICT}  {RGB}{255,  0,255}
\definecolor{colorNMIJ}  {RGB}{136,  0,136}
\definecolor{colorNIM}   {RGB}{  0, 64,255}

\begin{document}

\title{Optical to microwave clock frequency ratios with a nearly continuous strontium optical lattice clock}

\author{J\'er\^ome Lodewyck, S\l{}awomir Bilicki, Eva Bookjans, Jean-Luc Robyr, Chunyan Shi, Grégoire Vallet, Rodolphe Le Targat, Daniele Nicolodi, Yann Le Coq, Jocelyne Gu\'ena, Michel Abgrall, Peter Rosenbusch, S\'ebastien Bize}

\address{LNE-SYRTE, Observatoire de Paris, PSL Research University, CNRS, Sorbonne Universités, UPMC Univ. Paris 06, 61 Avenue de l'Observatoire, 75014 Paris, France}
\ead{jerome.lodewyck@obspm.fr}

\begin{abstract}
Optical lattice clocks are at the forefront of frequency metrology. Both the instability and systematic uncertainty of these clocks have been reported to be two orders of magnitude smaller than the best microwave clocks. For this reason, a redefinition of the SI second based on optical clocks seems possible in the near future. However, the operation of optical lattice clocks has not yet reached the reliability that microwave clocks have achieved so far. In this paper, we report on the operation of a strontium optical lattice clock that spans several weeks, with more than 80\% uptime. We make use of this long integration time to demonstrate a reproducible measurement of frequency ratios between the strontium clock transition and microwave Cs primary and Rb secondary frequency standards.
\end{abstract}

\ioptwocol

\section{Introduction}

Over the last years, optical lattice clocks (OLCs) have become the most stable frequency standards, with frequency instabilities as low as $2.2\times 10^{-16}$ in relative units after 1~s of integration~\cite{hinkley_atomic_2013,nicholson2015systematic}. More recently, by reducing the uncertainty on the blackbody radiation induced frequency shift, OLCs with accuracies in the $10^{-18}$ range have been demonstrated~\cite{bloom2014optical, ushijima2015cryogenic, nicholson2015systematic}.
On the other hand, their added complexity, mainly associated with the large number of laser sources operated at the same time, may \emph{a priori} hinder their ability to ever replace the well established cesium standards, including atomic fountain clocks. Indeed, practically realising the SI second requires that several primary standards are in almost continuous operation worldwide. While OLCs have been successfully demonstrated in many metrology laboratories, only sporadic operation was reported so far. In this context, linking optical clocks to international time scales requires extrapolation algorithms to fill the gaps in the operation of optical clocks~\cite{1347-4065-54-11-112401,ptbtimescale}. The fluctuations of the local oscillator realising the timescale -- so far an hydrogen maser -- during these gaps therefore remains the main source of uncertainty. 

To get rid of this uncertainty, it is necessary to operate reliable OLCs with a large time coverage. In this paper, we report on a step forward in this direction by demonstrating that an OLC using strontium atoms can be reliably operated over time periods of several weeks, with a time coverage larger than 80~\%.

We take advantage of these long integration times to compare one of our strontium clocks with three atomic fountains frequency standards with a statistical uncertainty below $10^{-16}$.

\section{Strontium optical lattice clocks}

\begin{figure}
	\includegraphics[width=\columnwidth]{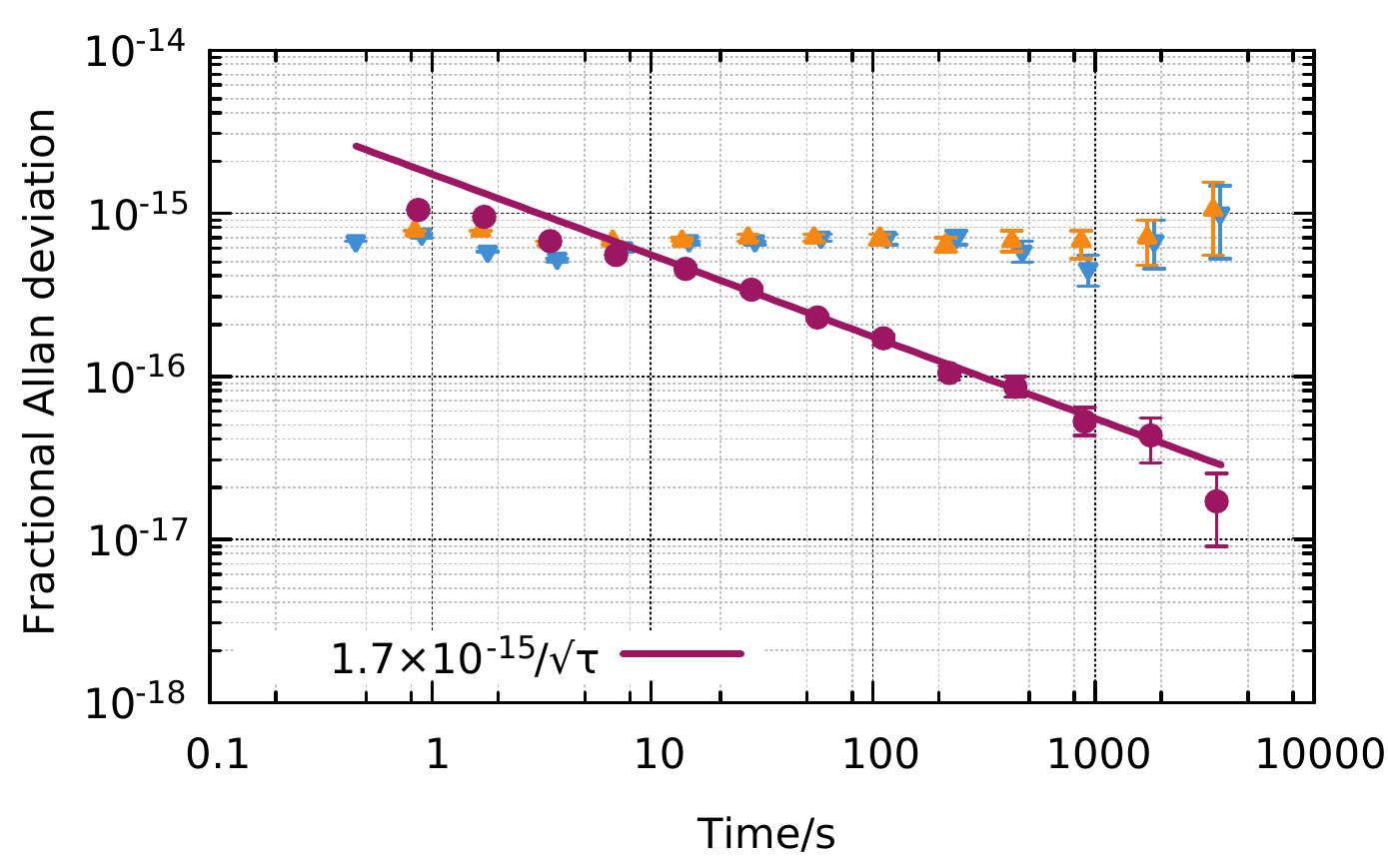}
	\caption{\label{fig:srstab}Frequency stability of two strontium optical lattice clocks. The blue $\color{Sr1color}\blacktriangledown$ (orange $\color{Sr2color}\blacktriangle$) show the Allan deviation of the frequency difference between the Sr1 (Sr2) lattice clock and the ultra-stable reference cavity. It exhibits the flicker noise level of the coatings of this cavity at $7\ex{-16}$. The purple $\color{SrDcolor}\bullet$ show the Allan deviation of the frequency difference between the two clocks when interrogated asynchronously. It reaches the mid-$10^{-17}$ range within one hour of integration.}
\end{figure}

\begin{table}
\begin{center}
\caption{\label{tab:sracc}Accuracy budget of the LNE-SYRTE Sr2 clock for Oct. 2014. The accuracy budget for Jun. 2015, found in~\cite{lisdat2015clock}, is equivalent, although a few terms are estimated with a slightly different uncertainty.}
\footnotesize
\begin{tabular}{@{}lll}
\br
Systematic effect & \begin{tabular}[x]{@{}c@{}}Correction\\/$10^{-18}$\end{tabular} & \begin{tabular}[x]{@{}c@{}}Uncertainty\\/$10^{-18}$\end{tabular}\\
\mr
		Blackbody radiation shift & $ 5208 $    & 20 \\
		Quadratic Zeeman shift     & $ 1317 $    & 12 \\
		Lattice light shift        & $ -30  $    & 20 \\
		Lattice spectrum           & $ 0 $       &  1 \\
		Density shift              & $ 0 $       &  8 \\
		Line pulling               & $ 0 $       & 20 \\
		Probe light shift          & $ 0.4 $     & 0.4\\
		AOM phase chirp            & $ -8 $      &  8 \\
		Servo error                & $ 0 $       &  3 \\
		Static charges             & $ 0 $       & 1.5\\
		Blackbody radiation oven  & $ 0 $       & 10 \\
		Background collisions      & $ 0 $       &  8 \\

\mr
		\textbf{Total}           & $6487.4$ & $41$\\
\br

\end{tabular}
\end{center}
\end{table}
\normalsize

Two optical lattice clocks with strontium atoms, Sr1 and Sr2, have been implemented at LNE-SYRTE. In~\cite{le_targat_experimental_2013}, we reported the first comparison between these two clocks with a total uncertainty lower than the accuracy of the best atomic fountains, thus confirming that these clocks actually overcame the systematic uncertainty of primary standards. The frequency stability obtained in a recent comparison between these two clocks is shown in Figure~\ref{fig:srstab}. The statistical resolution amounts to $10^{-15}$ at 1~s, and reaches the mid-$10^{-17}$ after one hour.

The up-to-date accuracy budget of the Sr2 clock used in this work is shown in Table~\ref{tab:sracc}.

The blackbody radiation (BBR) shift is estimated by continuously measuring the temperature of the vacuum chamber surrounding the atoms with 6 calibrated Pt100 resistor based thermometers. The main source of temperature inhomogeneity is the heat dissipated in the quadrupole coils used for the magneto-optical trap. In order to estimate the BBR shift, two sensors (T1 and T2) are placed close to these coils (hottest point) and two others (T3 and T4) are placed as far as possible from the coils (coldest point). Because we sample the temperature of the hottest and coldest points of the environment, the effective temperature experienced by the atoms is thus bounded by our temperature measurements. Without any further assumptions about the temperature distribution of the vacuum chamber, meaning that the effective temperature is equally likely to be anywhere between these bounds, the associated one standard deviation uncertainty is $1/\sqrt{12}$ of the spread of the temperature measurements~\cite{jcgm:2008:EMDG, Falke_2013}. In our case, this spread is 0.9~K, which corresponds to a one standard deviation uncertainty of 0.26~K. The frequency correction and uncertainty, reported in Table~\ref{tab:sracc}, are then deduced from the static and dynamic polarizabilities reported in~\cite{middelmann2012high}. The model described above to estimate the BBR shift is rather pessimistic. Indeed, we also monitor (thermometers T5 and T6) the temperature of two points in between the four other sensors, and measure temperatures close to the average of thermometer T1 to T4. In addition, the largest contribution to the BBR experienced by the atoms comes from two large windows placed on each side of the atomic sample, because they occupy a large solid angle, and because their emissivity is larger than that of the metallic parts of the chamber. We checked that the temperature of these windows is close to the temperature of T5 and T6. These measurements show that the probability distribution for the effective temperature is more likely to be around the average of the measured temperatures, rather than being uniform between the hottest and coldest temperature.

The Sr source is an oven heated at 873~K with an aperture of 5~mm diameter located 700~mm from the atoms. The  BBR directly hitting the atoms from the oven therefore yields a frequency shift of $10^{-18}$. However, this radiation is likely to be reflected by the vacuum environment, thus inducing a larger frequency shift. Accurately modelling these reflections is non trivial. We rather consider a simple model in which the atoms are surrounded by a sphere of radius 700~mm and emissivity 0.1 at ambient temperature. This sphere uniformly reflects the BBR emitted by a small BBR source on its surface with the same characteristics as the oven. In this model, we find that the BBR shift at the centre of the sphere is $10^{-17}$, which we use for the uncertainty of the BBR shift induced by the oven. This model is likely to give an overestimate of this uncertainty because most of the solid angle from the atoms has an emissivity larger than 0.1 (mainly glass and unpolished stainless steel and aluminium), and because the BBR indirectly emitted by the oven to the atoms is clamped by several thermal shields.

The quadratic Zeeman shift is derived from a real-time measurement of the Zeeman splitting between the $m_F = 9/2$ and $m_F=-9/2$ components and using the coefficient linking this splitting to the quadratic Zeeman effect published in~\cite{westergaard_lattice-induced_2011}. The uncertainty of the coefficient determines the uncertainty of the Zeeman effect.

The lattice light shift is estimated by measuring the clock frequency at different trap depths and fitting the data to a parabola~\cite{westergaard_lattice-induced_2011, PhysRevA.92.012516}. The trap depths are varied between
$50~E_R$ to $1000~E_R$ -- $E_R$ being the recoil energy associated with the absorption of a lattice photon. The curvature of this parabola arises from hyperpolarizability effects~\cite{le2012comparison}, and is resolved in a single measurement when comparing the clock frequency for different trap depths up to 1000~$E_R$. We can quantitatively reproduce this measurement with our two strontium clocks, with different laser sources for the lattice light, and with different atomic densities. In nominal operation of the clocks, the trap depth is of typically $70~E_R$, and the light shift correction and uncertainty are deduced from the parabolic fit. The measurements reported in this paper were conducted either with a lattice wavelength detuned by 10~MHz from the magic wavelength (as reported in Table~\ref{tab:sracc}), or tuned to the magic wavelength.

On top of this regular light shift, we observed a frequency shift related to the incoherent background spectrum of semiconductor sources~\cite{le2012comparison}. This shift was observed to vary over time and from chip to chip. For the measurements reported in this paper, we instead used a titanium-sapphire (TiSa) laser source (M SQUARED SolsTiS). While we expect that the incoherent background of a TiSa laser is lower than that of semiconductor sources, a quantitative study is required to upper-bound the associated systematic uncertainty. For this, we measured the total power and spectral distribution of the spontaneous emission transmitted from the laser to the atoms when the round trip in the TiSa laser cavity was blocked. Because the amount of spontaneous emission emitted by the pumped crystal is larger when the laser cavity is not lasing, this configuration gives a worst-case estimate of the associated induced frequency shift on the atoms. We measured a total power of spontaneous emission 89 dB below the power of the lasing TiSa. Given the spectral distribution of this light, spanning up to 100~nm away from the magic wavelength, the corresponding frequency shift is lower than $10^{-18}$ for a trap depth of 100~$E_R$.

We also investigated possible non linear effects in the TiSa laser cavity that are not visible when the laser emission is blocked, such as Raman scattering. While these effects are expected to be corrected by a standard extrapolation of the clock frequency at zero trap depth, they may fluctuate in power or spectral distribution during the clock operation and lead to an instability or a frequency shift. To estimate this shift, we compared the clock frequency when the TiSa is nominally operated at full power (2 W output) and when it is operated at reduced power (0.9 W output, by reducing the pump power), while keeping the trap depth constant at about 1500~$E_R$ by attenuating the laser light between the TiSa and the atoms. Any relevant nonlinearity in the laser cavity would translate into a frequency shift between these two configurations, which we repeatedly measured to be below $5\times 10^{-17}$. The average of these measurements, extrapolated at the nominal working point of the clock (TiSa at full power and a trap depth below $100~E_R$), yields a maximal frequency shift of $7 \ex{-18}$.

The density shift is measured by operating the clock alternatively at the nominal number of atoms (roughly 3 to 5 atoms per site), and with a reduced number of atoms (close to 1 atom per site). No frequency shift is observed between the two configurations with a statistical uncertainty of $8 \ex{-18}$.

The line pulling shift results from the possible presence of extra lines nearby the clock resonance, arising for instance from imperfect optical pumping. The presence of such lines distorts  the unperturbed resonance, shifting the position of its maximum. This effect is smaller for narrower resonances because a narrow resonance has a sharper, hence less perturbed maximum. Also, if nearby resonances are narrower, their pedestal is smaller and therefore less perturbing. To estimate an upper bound on this effect, we alternatively probe the clock transition with different probing time (100~ms and the nominal 200~ms duration), hence different linewidths, resulting in different pulling. We observe no effect with a statistical uncertainty of $2\ex{-17}$.

The probe light shift is estimated from a theoretical calculation of the polarizabilities of the clock levels at the wavelength of the clock laser (3~mHz/(W/m$^2$)) and the light intensity ($I = 50$~mW/m$^2$) deduced from the fact that we probe the clock transition with a $\pi$ pulse.

The AOM phase chirp is measured by an heterodyne optical interferometer comparing the phase of the pulsed light to the phase of a non-pulsed reference arm. We observe that the effect can vary with the alignment of the optical beam, and therefore we attribute a large uncertainty on the effect. For the parameters used to induce a $\pi$ pulse on the atomic transition, the effect is thus estimated to $8 (8) \ex{-18}$ (Oct. 2014). For higher accuracy, we have the possibility to reduce the effect by more than one order of magnitude by increasing the optical power and consequently reducing the RF power feeding the AOM (Jun. 2015).

The frequency drift of the ultra-stable reference cavity is pre-compensated using a DDS based de-drifting system using a third order polynomial drift. The three coefficients of this polynomial are updated every 15 minutes by fitting the time dependence of the frequency correction of the clock. The residual drift rate is lower than 1~mHz/s, which corresponds to a servo systematic uncertainty below $3\ex{-18}$.

The estimation of the frequency shift due to static charges is reported in~\cite{lodewyck2012observation}, and background collisions in~\cite{PhysRevLett.110.180802}.

The total systematic frequency uncertainty for Sr2 is $4.1\ex{-17}$.

\section{Cesium and rubidium atomic fountains}

Among the ensemble of atomic fountains developed at LNE-SYRTE~\cite{guena2012}, two are involved in this work: FO1, operated with cesium atoms, and the dual fountain FO2 using simultaneously rubidium and cesium atoms, FO2-Cs and FO2-Rb~\cite{guena2010}.

\begin{figure}
	\includegraphics[width=\columnwidth]{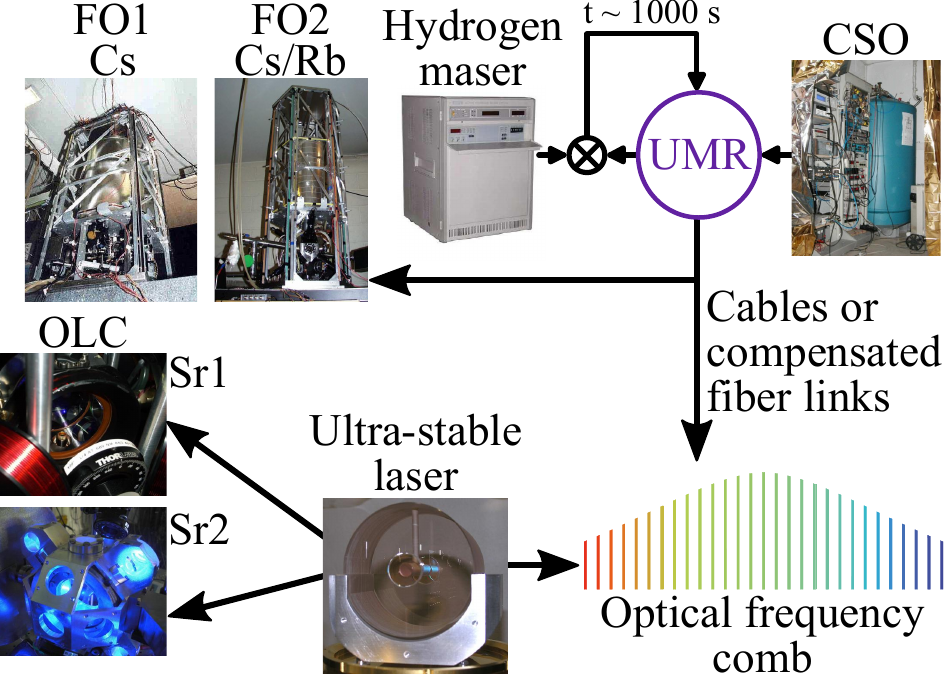}
	\caption{\label{fig:clock_ensemble}Simplified scheme of the LNE-SYRTE atomic clock ensemble involved in this work. UMR: Ultra-stable Microwave Reference, CSO: Cryogenic Sapphire Oscillator, OLC: Optical Lattice Clock.}
\end{figure}

As shown in Figure~\ref{fig:clock_ensemble}, FO1, FO2-Cs, and FO2-Rb measure the frequency of the same ultra-stable microwave frequency reference (UMR) which is a signal at 11.98~GHz based on a cryogenic sapphire oscillator (CSO), and is phase locked to a hydrogen maser with a time constant of the order of 1000~s. The UMR signal thus exhibits the low phase noise of the CSO at small time scales, which corresponds to a stability of the order of $10^{-15}$ at 1~s, and then converges to the slowly drifting H-maser frequency. The UMR signal is also distributed via a fibre link to the frequency comb (section~\ref{sec:femto}).

FO1, FO2-Cs and FO2-Rb operate with optical molasses loaded from a 2D magneto-optical trap. Thanks to the use of the UMR, the stability of the fountains is quantum projection noise  limited with a rather high number of atoms detected (a few $10^6$). During the measurement campaigns reported in this work, the stability at 1 s varies between  3 to 9$\times 10^{-14}\tau^{-1/2}$ depending upon the fountain and on deliberately varying the atom number.

The strategies to reduce and evaluate the main frequency shifts and associated uncertainties, in quasi real-time for the main shifts, are described in ~\cite{guena2011, guena2012, abgrall2015461, guena2014}. Last published accuracy budgets can be found in ~\cite{guena2012}. During the measurements reported in this work, the accuracy of the fountains reached $3.6\times 10^{-16}$, $2.4\times 10^{-16}$ and $2.9\times 10^{-16}$ for FO1, FO2-Cs and FO2-Rb, respectively.

The operation of FO1 and FO2 with Cs and Rb simultaneously has been nearly continuous for the last 8 years, regularly providing calibration reports to the BIPM to participate in the steering of TAI and in the definition of the SI second.
The process of using a secondary representation of the second in the steering of TAI has been initiated with FO2-Rb~\cite{guena2014} which also regularly provides calibration data to the BIPM (SYRTE-FORb in BIPM \emph{Circular T}). This is a useful step in view of a future redefinition of the SI second based on optical clocks. In this work we present the first comparison of FO2-Rb with one Sr clock with an uncertainty similar to Sr/Cs comparisons. One interest is that the Sr/Rb optical-to-microwave frequency ratio presents new sensitivities to possible variations of the fundamental constants.

During the three-week comparison campaign reported in section 6, the total uptimes of the fountains were 89\%, 93\% and 94\% for FO1, FO2-Cs and FO2-Rb respectively.

\section{Frequency comb and microwave link}
\label{sec:femto}

The SYRTE optical frequency comb dedicated to operational measurements provides the frequency ratio between the UMR measured by the fountains and the ultra-stable optical reference (Sr clock laser at 698~nm) measured by the OLC (Figure~\ref{fig:clock_ensemble}). 
The comb is generated by a pulsed erbium-doped fibre laser, with a repetition rate $f_{\mathrm{rep}} \simeq 250$~MHz. The carrier offset frequency $f_0$ is mixed out of the beat note between the comb and an ultra-stable IR laser at 1542 nm, and the resulting signal is phase locked to a synthesiser at 880~MHz, so that the spectral purity of the IR laser (flicker floor at $5\times10^{-16}$) is transferred to $f_{\mathrm{rep}}$. With this approach, the virtual comb $N \times f_{\mathrm{rep}} \pm 880$~MHz is in the narrow linewidth regime.
The feedback loop acts both on a fast actuator (electro-optic modulator controlling the optical length of the fibre laser cavity) and a slow actuator (piezoelectric actuator controlling the geometric length of the cavity), resulting in a bandwidth of 900~kHz~\cite{Zhang2012}.

The comb is compared to the UMR via a compensated microwave link connecting the UMR lab and the optical frequencies lab. A signal at 8.985~GHz derived from the UMR ($3/4$ of $11.98$~GHz) is used to modulate the amplitude of a 1.5~$\mu$m laser diode injected in a 200~meter fibre.
At the output, a small part of the light is coupled back into the fibre, and sent back towards the input where it is demodulated. The phase difference between the source and the demodulated microwave is locked to 0 (modulo $2\pi$) via a feedback to a fibre stretcher and to a fibre heater. Therefore the phase fluctuations introduced by the propagation in the fibre are compensated.
The main part of the signal is demodulated to produce a ``maser-exact'' signal at 8.985~GHz. The harmonics of the comb's repetition rate are detected by a Discovery Semiconductors high speed photo-diode DSC40S, the 36\textsuperscript{th} harmonics around 9~GHz is selected by a cavity filter (rejecting the next harmonics by at least 40~dB) and amplified by an AML612L2201 amplifier. Lastly, the frequency of the 36\textsuperscript{th} harmonics is counted against the 8.985~GHz signal.

The comb light is amplified by a dedicated erbium-doped fibre amplifier and goes through a highly non-linear fibre whose length was adjusted to favour four-wave mixing around 1396~nm. It is then frequency doubled in a periodically poled lithium niobate crystal, in order to generate teeth around~698 nm, thus allowing us to generate a beat note between the comb and 1~mW of strontium clock laser light. Light from the strontium ultra-stable reference cavity is transferred to the frequency comb via a 40~m fibre, equipped a with phase noise cancellation system that actively suppresses length fluctuations of the optical
After mixing out $2f_0$, the beat note has a linewidth of about 1~Hz, with a signal-to-noise ratio of 35~dB in a resolution bandwidth of 1~kHz. The signal is filtered in a bandwidth of 1~kHz which is automatically recentred on the average frequency of the beat note every 30~minutes in order to accommodate for the drifts of the lasers. The signal is counted by a dead-time-free counter also referenced to the UMR.

In summary, the frequency comb provides the dimensionless frequency ratio between the strontium ultra-stable cavity and the UMR, with measurement intervals of 1 second synchronised to the LNE-SYRTE local time scale UTC(OP) with a delay lower than $10~\mu$s. The inaccuracy of this particular configuration of the operational comb was checked to be at most at the $1 \times 10^{-16}$ level by synchronously measuring the Sr clock laser with a comb based on a titanium-sapphire laser.

\section{Operability of a strontium clock}

\begin{figure*}
\begin{center}
	\includegraphics[width=\textwidth]{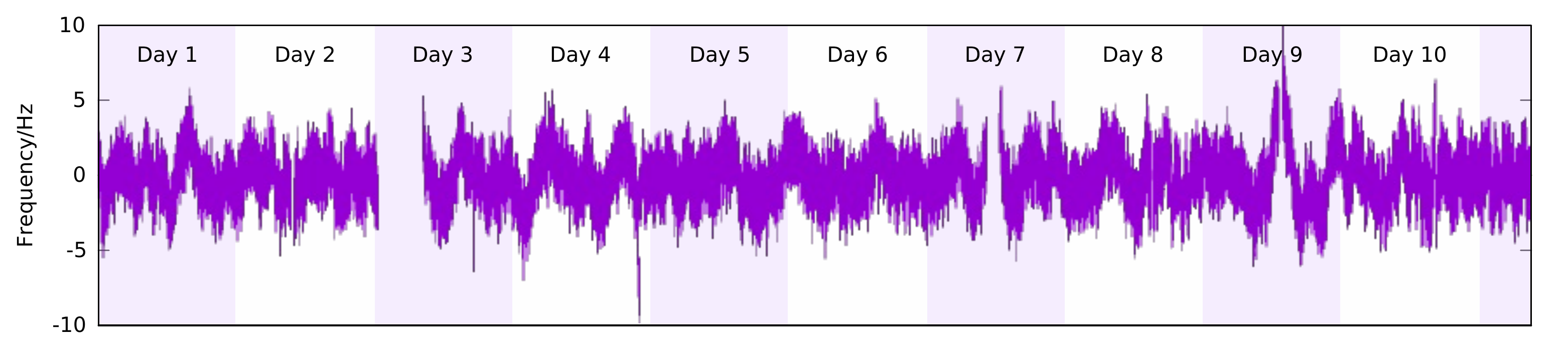}\\
	\includegraphics[width=\textwidth]{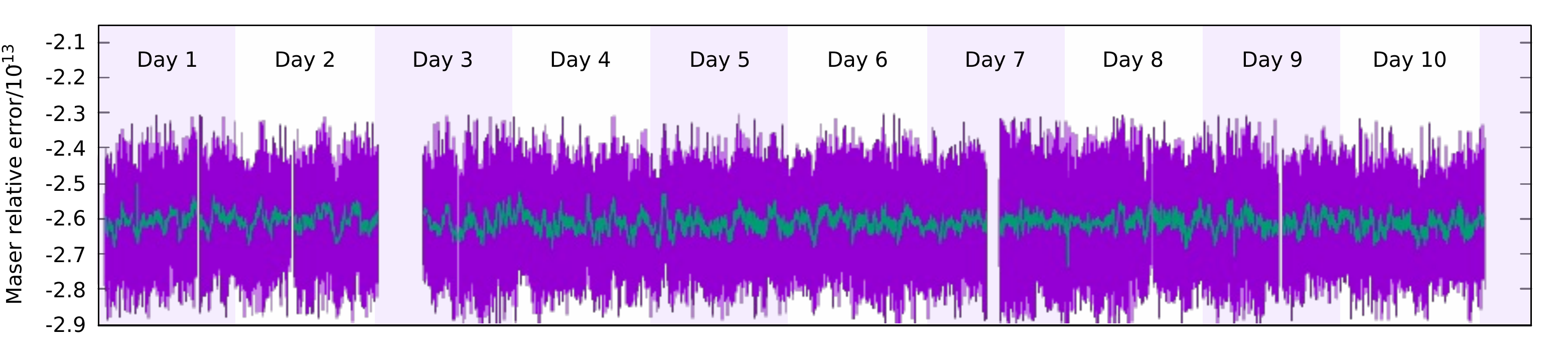}
	\caption{\label{fig:temporal}Top: difference between the clock frequency of the Sr2 clock and the resonance frequency of the ultra-stable reference cavity. The drift of the cavity is removed by fitting with polynomial up to 5\textsuperscript{th} order over successive intervals of 1 or 2 days. The residual fluctuations come from the thermal noise of the coatings of the ultra-stable cavity. Bottom: hydrogen maser relative frequency error, referenced against the Sr atoms via the frequency comb. The data span over ten days, from Oct. 24\textsuperscript{th}, 2014 to Nov. 2\textsuperscript{nd}, 2014. The green curve shows the average of these data over 100~s, showing that the strontium clock can efficiently average the maser noise over short time scales.}
\end{center}
\end{figure*}

The two strontium clocks Sr1 and Sr2 have been designed in order to be able to run mostly autonomously over long periods. All the laser systems have been equipped with relocking systems based on digital microcontrollers driven by algorithms that can detect and correct most common failures in the course of the clock operation. To demonstrate the efficiency of these systems, and to show that the Sr clocks are reaching an operational working point, we operated the Sr2 clock for 10 consecutive days in October 2014, and 3 consecutive weeks in June 2015. During these campaigns, the clock was fully unattended at night, while limited manual tuning was required during day time.

During the first campaign, Sr2 was operated with a cycle time of 0.95~s, and 93.8\% (see Figure~\ref{fig:temporal}) of the total cycles issued valid data points. Among these data, we found an interval of 90~hours with a 99.7\% uptime, comprising a night-time interval of 7~hours with a 100\% uptime. The total uptime including the frequency comb is 93\% over ten days, illustrating that the frequency comb does not limit the reliability of optical clocks. During the second three weeks campaign, the total uptime is 83\%.

\section{Measurement of frequency ratios}

\begin{figure}
\begin{center}
	\includegraphics[width=\columnwidth]{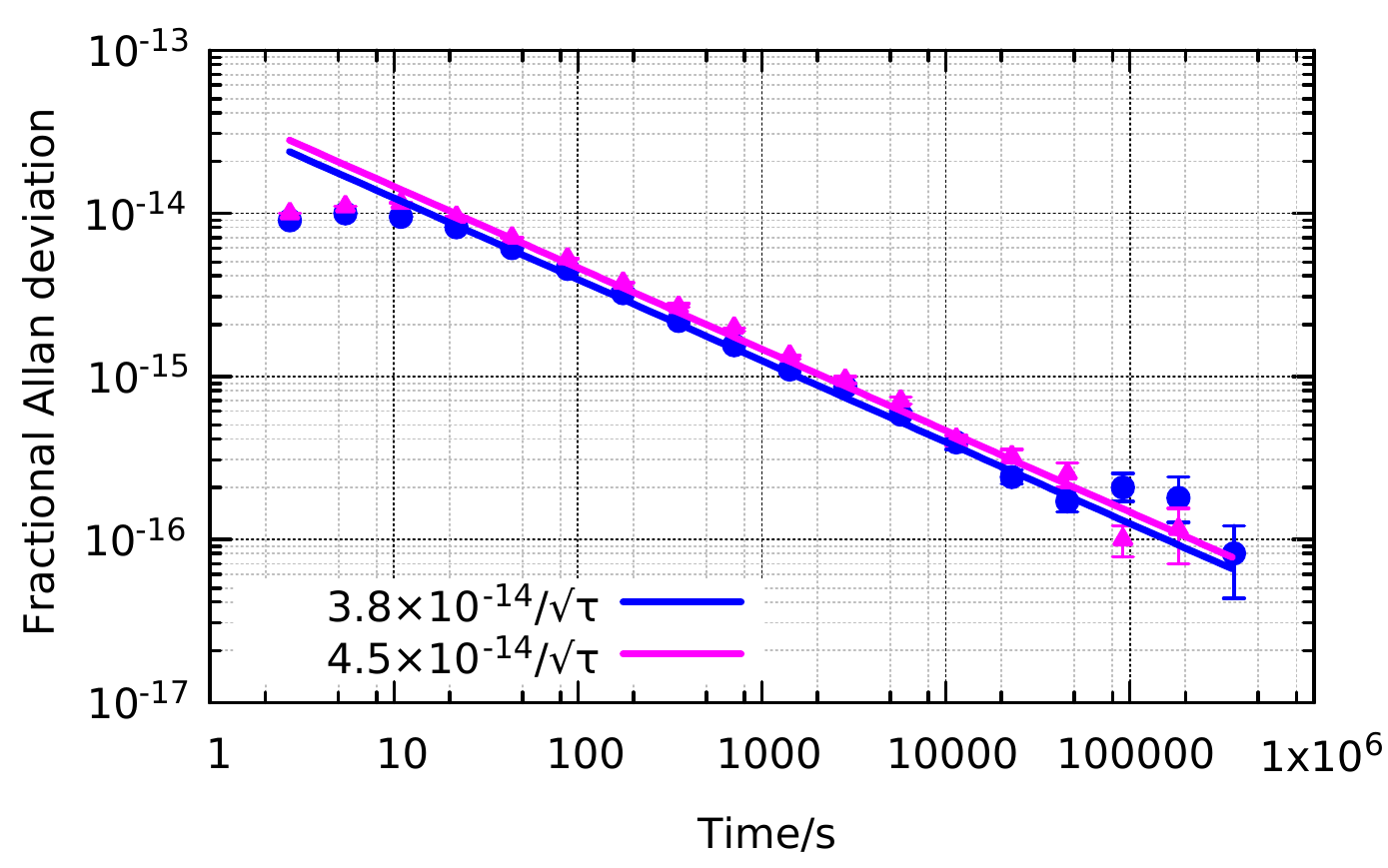}
	\caption{\label{fig:allanFO}Fractional Allan deviation of the frequency comparison between the Sr2 clock and the FO2 dual fountain using Cs (blue $\color{blue}\bullet$) simultaneously with Rb (pink $\color{magenta}\blacktriangle$) over ten days (October 2014). The stability is limited by the quantum projection noise of the fountain, and reaches a statistical resolution below $10^{-16}$ after a few days of measurement.}
\end{center}
\end{figure}

\begin{figure*}
\begin{center}
	\includegraphics[width=\columnwidth]{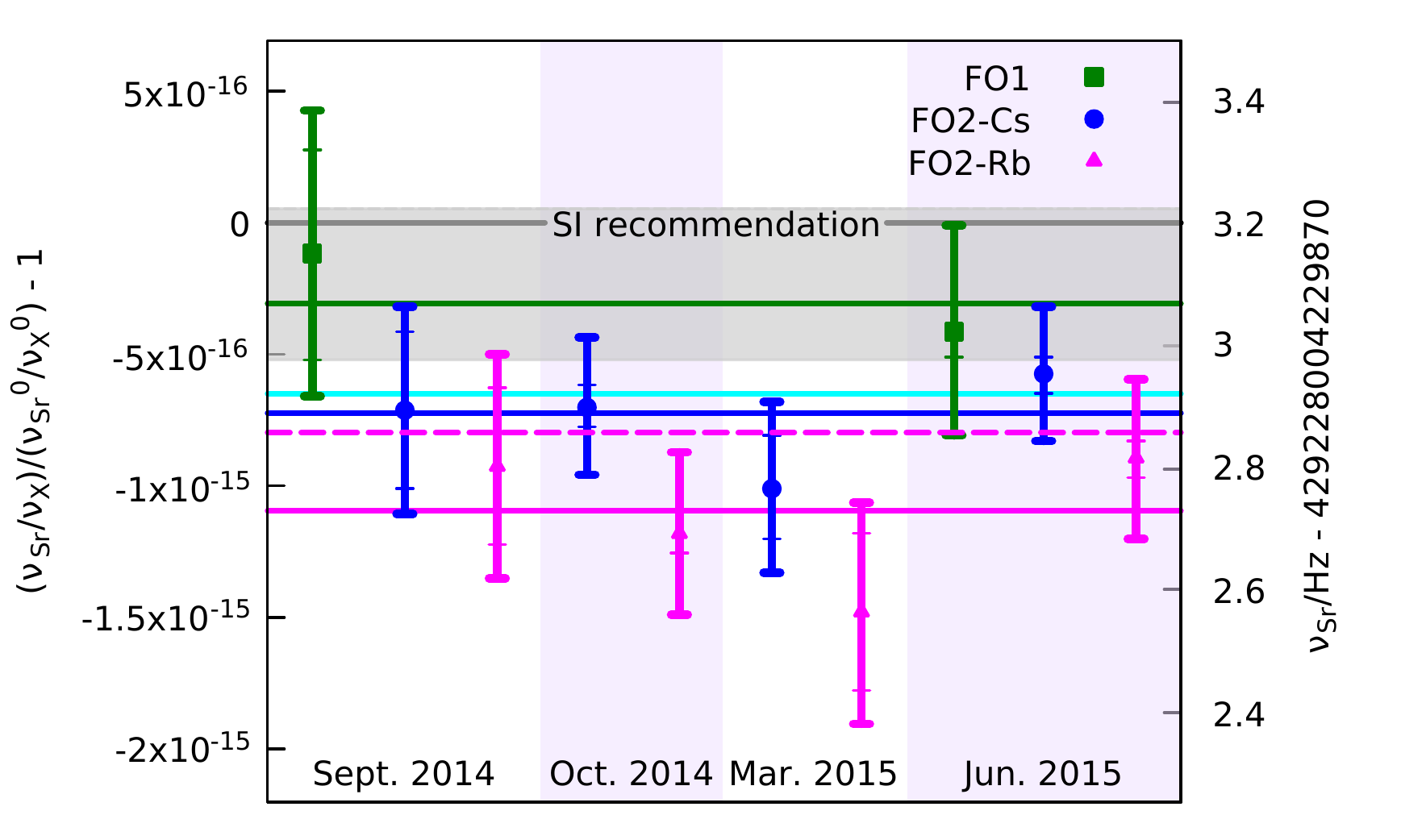}\hfill
	\includegraphics[width=\columnwidth]{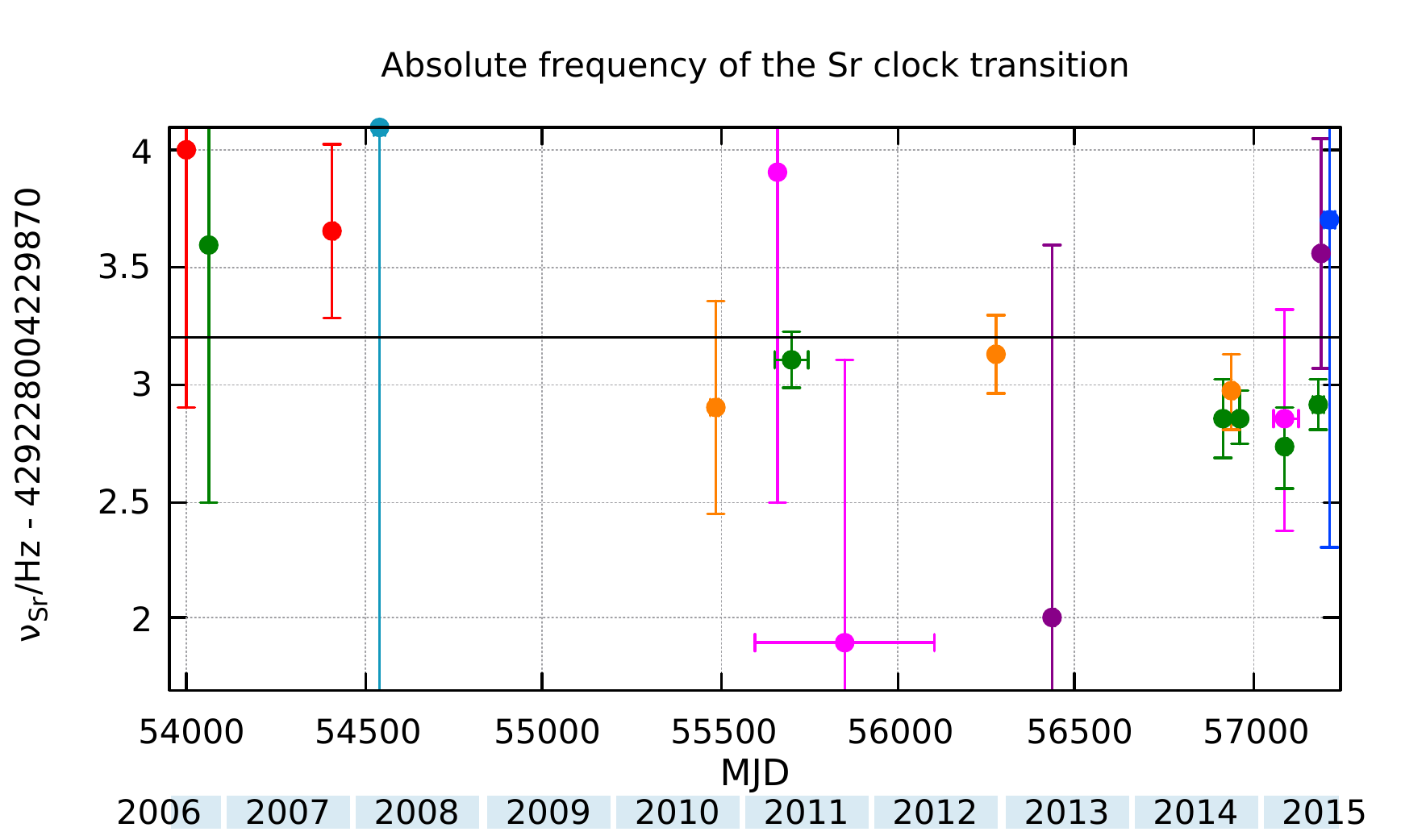}
	\caption{\label{fig:SrVsFO}Left: Frequency ratio between strontium and cesium or rubidium clock transitions. For each point, the statistical uncertainty is represented with a thin error bar, and the total uncertainty with a thick error bar. The left axis shows the relative offset between the measured frequency ratio and the frequency ratio given the secondary representations of the SI second for strontium ($\nu_\textrm{\tiny Sr}^0 = 429\,228\,004\,229\,873.2$~Hz, gray line on the graph) and rubidium ($\nu_\textrm{\tiny Rb}^0 = 6\,834\,682\,610.904\,310$~Hz) recommended by the CIPM~\cite{CIPM}. For the ratio between strontium and cesium, the right axis translates this offset to the absolute frequency of the strontium clock transition in the SI. The colour lines indicate the per-fountain average reported in Table~\ref{tab:freqratio}. To illustrate the reproducibility of the Rb/Cs frequency ratio with FO2, the dashed magenta line represents the frequency ratio between Sr and Rb with respect to the Rb frequency measured at LNE-SYRTE ($\nu_\textrm{\tiny Rb}^\textrm{\tiny SYRTE} = 6\,834\,682\,610.904\,312$~Hz) instead of the recommended value $\nu_\textrm{\tiny Rb}^0$. The cyan line is the average over the two Cs fountains. The grey area represents the uncertainty of the  Sr/Cs frequency ratio we previously reported in~\cite{le_targat_experimental_2013}. Right: the points for Cs of the left figure with the smallest statistical uncertainty are compared with similar measurement worldwide since 2006 (green $\color{colorSYRTE}\bullet$: LNE-SYRTE~\cite{baillard2008optical,le_targat_experimental_2013}, red $\color{colorJILA}\bullet$: JILA~\cite{0026-1394-45-5-008,PhysRevLett.100.140801}, light blue $\color{colorTokyo}\bullet$: Tokyo University~\cite{Hong:09}, orange $\color{colorPTB}\bullet$: PTB~\cite{0026-1394-48-5-022, Falke_2013, ptbtimescale}, pink $\color{colorNICT}\bullet$: NICT~\cite{Hong:09,1882-0786-5-2-022701,Matsubara:12,1347-4065-54-11-112401}, purple $\color{colorNMIJ}\bullet$: NMIJ~\cite{1882-0786-7-1-012401,doi:10.7566/JPSJ.84.115002}, blue $\color{colorNIM}\bullet$: NIM~\cite{0256-307X-32-9-090601}.}
\end{center}
\end{figure*}

\begin{table*}
\begin{center}
\caption{\label{tab:freqratio}Frequency ratios $\frac{\nu_\textrm{\tiny Sr}}{\nu_\textrm{\tiny X}}$ between the Sr2 OLC and the atomic fountains. The $X$ subscript stands for Cs or Rb. The third column is the relative offset between these ratios and the ratios between the recommended value of the frequency for Rb and Sr~\cite{CIPM}, and the definition for Cs, indicated by a $^0$ superscript. In the case of Cs, the forth column translates this value into the absolute frequency of strontium. The total fractional uncertainty on each representation of the frequency ratios is given in parenthesis. The last column is the reduced $\chi^2$ of the average fit taking into account the total uncertainty (applies to FO2-Cs and FO2-Rb, for which we have the largest data set). It is lower than one, indicating that the systematic frequency shift is stable over time. The data of columns 3 and 4 are shown in Figure~\ref{fig:SrVsFO} (left).}
\footnotesize
\begin{tabular}{@{}lcccc}
\br
Fountain & $\frac{\nu_\textrm{\tiny Sr}}{\nu_\textrm{\tiny X}}$ & $\frac{\nu_\textrm{\tiny Sr}}{\nu_\textrm{\tiny X}} \frac{\nu^0_\textrm{\tiny X}}{\nu^0_\textrm{\tiny Sr}} - 1$ & $\nu_\textrm{Sr}$/Hz & $\chi^2$\\
\mr
		FO1      & $46\,692.613\,711\,630\,600 (18)$ & $ -3.0 (39) \ex{-16} $ & $429\,228\,004\,229\,873.07 (17)$ &  \\
		FO2-Cs   & $46\,692.613\,711\,630\,580 (13)$ & $ -7.2 (28) \ex{-16} $ & $429\,228\,004\,229\,872.89 (12)$ &  0.6 \\
		Cs total & $46\,692.613\,711\,630\,583 (13)$ & $ -6.5 (28) \ex{-16} $ & $429\,228\,004\,229\,872.92 (12)$ &  \\
		FO2-Rb   & $62\,801.453\,800\,512\,435 (21)$ & $ -10.9 (33) \ex{-16} $ &      &  0.5 \\ 
\mr
\end{tabular}
\end{center}
\end{table*}
\normalsize

During the two measurement campaigns described in the previous section, we measured the frequency ratios between the strontium clock transition and microwave clock transitions by comparing the Sr2 clock to the FO2-Cs, FO2-Rb, and the FO1 Cs fountains. The frequency stability of Sr2 vs. FO2-Cs and FO2-Rb is shown in Figure~\ref{fig:allanFO}. One can see that given the stability of the microwave standards, long integration times are indeed required in order to reach a statistical resolution below $10^{-16}$. Figure~\ref{fig:SrVsFO} shows the corresponding frequency ratios measured during the October 2014 and June 2015 campaigns. These measurements both feature a negligible statistical uncertainty. Figure~\ref{fig:SrVsFO} also shows two other measurements lasting a few hours. Table~\ref{tab:freqratio} lists the average of these four measurements of Sr against Cs and Rb.

The total uncertainty of these ratios comprises the statistical uncertainty, the systematic uncertainty of the clocks described above, the uncertainty of the fibre link between the microwave and optical clocks of $10^{-16}$, the uncertainty of the frequency comb of $10^{-16}$, as well as the uncertainty on the redshift correction. A recent levelling campaign enabled us to determine the height difference between all the clocks of LNE-SYRTE with an uncertainty of at most 2~cm. In the case of the atomic fountains, this levelling takes into account the relativistic frequency shifts over the trajectory of the moving atoms (second order Doppler effect and gravitational redshift). Hence, the effective height difference (resp. relativistic corrections on the frequency ratio) between Sr2 and FO2-Cs is $-4.29 (2)$~m (resp. $4.68 (2)\times 10^{-16}$), between Sr2 and FO2-Rb is $-4.21 (2)$~m (resp. $4.59 (2)\times 10^{-16}$), and between Sr2 and FO1 is $-7.529 (20)$~m (resp. $8.22 (2)\times 10^{-16}$).

The ratio between Sr and Cs, yielding the absolute frequency of the Sr clock transition in the SI unit system, has been extensively measured over time with increasing resolution, at LNE-SYRTE, as well as in other laboratories. The average of the measurements presented in this paper is:
\begin{equation}
	\nu_\textrm{Sr} = 429\,228\,004\,229\,872.92 (12) ~\textrm{Hz},
\end{equation}
all uncertainties combined. This result differs by $-4.2\ex{-16}$, \emph{i.e.} one combined standard uncertainty, from our last measurement reported in~\cite{le_targat_experimental_2013}~\footnote{In the current paper, we use a refined estimate of the height difference between the clocks. Using this new estimate, the Sr/Cs frequency ratio published in~\cite{le_targat_experimental_2013} would be higher by $6\ex{-17}$, thus bringing the discrepancy to 1.14 standard uncertainty.}. This value is in very good agreement with other international high resolution measurements of the Sr/Cs frequency ratio (Figure~\ref{fig:SrVsFO}, right).

Additionally, this work reports on the first measurement of the frequency ratio between Sr and Rb. This measurement reaches the same uncertainty as the Sr/Cs frequency ratio reported in this work (Table~\ref{tab:freqratio}). Figure~\ref{fig:SrVsFO} and the third column of table~\ref{tab:freqratio} show that for each of our measurements, the Sr/Rb ratio is in very good agreement with the ratio between Sr/Cs ratio, given the recommended value of the $^{87}$Rb secondary representation of the second, which is mainly determined from long series of comparisons between Cs and Rb in the FO2 fountain~\cite{guena2014}. Repeating such interspecies frequency measurements over time will contribute to the tracking of possible variations of fundamental constants.

\section{Conclusion}

In this paper, we have shown that OLCs can be reliably operated over long periods with a time coverage larger than 80\%. This high reliability was obtained by implementing monitoring and control hardware and software on the clock subsystems. Besides enabling measurement of optical to microwave frequency ratios with a high statistical resolution, as presented here, the time coverage is compatible with the required performances to contribute to international time scales. Indeed, reference~\cite{ptbtimescale} shows that taking advantage of the high statistical resolution achieved by optical clocks to steer the frequency of an hydrogen maser with a faster bandwidth than what is achievable with microwave clocks can already improve existing atomic timescales, even with significant gaps in the operation of optical clocks. Applying this steering with the quasi-continuous data presented in this paper would shrink down the main uncertainty coming from the extrapolation of the maser behaviour during interruptions of the optical clock. However, maintaining a practical realisation of a timescale with optical clocks would require to make sure that hours long interruptions in the optical clocks would not mask a rapid frequency jump of the hydrogen maser, for instance by having a set of two optical clocks.

Ultimately, a fully optical timescale could be realised using the building blocks presented in this paper, comprising a reliable photonic oscillator and a femtosecond frequency comb, almost continuously measured by a set of operational lattice clocks. This work therefore opens the way to build an operational architecture with optical clocks that will be able to replace the microwave based frequency chain currently in use for time and frequency standards.

\section*{Acknowledgements}

We acknowledge funding support from Centre National d'\'Etudes Spatiales (CNES), Conseil R\'egional \^Ile-de-France (DIM Nano'K), Agence Nationale de la Recherche (Labex First-TF ANR-10-LABX-48-01), and the European Metrology Research Programme (EXL-01 QESOCAS and SIB-55 ITOC). The EMRP is jointly funded by the EMRP participating countries within EURAMET and the European Union. We thank Pacôme Delva for supporting the height determination of the clocks.

\section*{References}

\bibliographystyle{unsrt}
\bibliography{operational_sr}

\begin{thebibliography}{10}

\bibitem{hinkley_atomic_2013}
N.~Hinkley, J.~A. Sherman, N.~B. Phillips, M.~Schioppo, N.~D. Lemke, K.~Beloy,
  M.~Pizzocaro, C.~W. Oates, and A.~D. Ludlow.
\newblock An atomic clock with {$10^{-18}$} instability.
\newblock {\em Science}, 341(6151):1215--1218, 2013.

\bibitem{nicholson2015systematic}
T.L. Nicholson, S.L. Campbell, R.B. Hutson, G.E. Marti, B.J. Bloom, R.L.
  McNally, W.~Zhang, M.D. Barrett, M.S. Safronova, G.F. Strouse, et~al.
\newblock Systematic evaluation of an atomic clock at {$2\times 10^{-18}$}
  total uncertainty.
\newblock {\em Nature communications}, 6:6896, 2015.

\bibitem{bloom2014optical}
B.~J. Bloom, T.~L. Nicholson, J.~R. Williams, S.~L. Campbell, M.~Bishof,
  X.~Zhang, W.~Zhang, S.~L. Bromley, and J.~Ye.
\newblock An optical lattice clock with accuracy and stability at the
  {$10^{-18}$} level.
\newblock {\em Nature}, 506:12941, 2014.

\bibitem{ushijima2015cryogenic}
I.~Ushijima, M.~Takamoto, M.~Das, T.~Ohkubo, and H.~Katori.
\newblock Cryogenic optical lattice clocks.
\newblock {\em Nature Photonics}, 9(3):185--189, 2015.

\bibitem{1347-4065-54-11-112401}
H.~Hachisu and T.~Ido.
\newblock Intermittent optical frequency measurements to reduce the dead time
  uncertainty of frequency link.
\newblock {\em Japanese Journal of Applied Physics}, 54(11):112401, 2015.

\bibitem{ptbtimescale}
S.~Dörscher S. Häfner V. Gerginov S. Weyers B. Lipphardt F. Riehle U. Sterr
  C.~Lisdat C.~Grebing, A. Al-Masoudi.
\newblock Realization of a time-scale with an optical clock.
\newblock {\em arXiv:1511.03888}, 2015.

\bibitem{lisdat2015clock}
C.~Lisdat, G.~Grosche, N.~Quintin, C.~Shi, S.~Raupach, C.~Grebing, D.~Nicolodi,
  F.~Stefani, A.~Al-Masoudi, S.~D{\"o}rscher, et~al.
\newblock A clock network for geodesy and fundamental science.
\newblock {\em arXiv preprint arXiv:1511.07735}, 2015.

\bibitem{le_targat_experimental_2013}
R.~Le~Targat, L.~Lorini, Y.~Le~Coq, M.~Zawada, J.~Guéna, M.~Abgrall, M.~Gurov,
  P.~Rosenbusch, D.~G. Rovera, B.~Nagórny, R.~Gartman, P.~G. Westergaard,
  M.~E. Tobar, M.~Lours, G.~Santarelli, A.~Clairon, S.~Bize, P.~Laurent,
  P.~Lemonde, and J.~Lodewyck.
\newblock Experimental realization of an optical second with strontium lattice
  clocks.
\newblock {\em Nature Communications}, 4:2109, 2013.

\bibitem{jcgm:2008:EMDG}
Joint~Committee for Guides~in Metrology.
\newblock Jcgm 100: Evaluation of measurement data - guide to the expression of
  uncertainty in measurement.
\newblock Technical report, JCGM, 2008.

\bibitem{Falke_2013}
S.~Falke, N.~Lemke, C.~Grebing, B.~Lipphardt, S.~Weyers, V.~Gerginov,
  N.~Huntemann, C.~Hagemann, A.~Al-Masoudi, S.~Häfner, S.~Vogt, U.~Sterr, and
  C.~Lisdat.
\newblock A strontium lattice clock with $3\times10^{-17}$ inaccuracy and its
  frequency.
\newblock {\em New Journal of Physics}, 16(7):073023, 2014.

\bibitem{middelmann2012high}
T.~Middelmann, S.~Falke, C.~Lisdat, and U.~Sterr.
\newblock High accuracy correction of blackbody radiation shift in an optical
  lattice clock.
\newblock {\em Physical review letters}, 109(26):263004, 2012.

\bibitem{westergaard_lattice-induced_2011}
P.~G. Westergaard, J.~Lodewyck, L.~Lorini, A.~Lecallier, E.~A. Burt, M.~Zawada,
  J.~Millo, and P.~Lemonde.
\newblock Lattice-induced frequency shifts in {S}r optical lattice clocks at
  the {$10^{-17}$} level.
\newblock {\em Physical Review Letters}, 106(21):210801, 2011.

\bibitem{PhysRevA.92.012516}
C.~Shi, J.-L. Robyr, U.~Eismann, M.~Zawada, L.~Lorini, R.~Le~Targat, and
  J.~Lodewyck.
\newblock Polarizabilities of the $^{87}\mathrm{Sr}$ clock transition.
\newblock {\em Phys. Rev. A}, 92:012516, 2015.

\bibitem{le2012comparison}
R.~Le~Targat, L.~Lorini, M.~Gurov, M.~Zawada, R.~Gartman, B.~Nag{\'o}rny,
  P.~Lemonde, and J.~Lodewyck.
\newblock Comparison of two strontium optical lattice clocks in agreement at
  the $10^{-16}$ level.
\newblock In {\em European Frequency and Time Forum (EFTF)}, pages 19--22,
  2012.

\bibitem{lodewyck2012observation}
J.~Lodewyck, M.~Zawada, L.~Lorini, M.~Gurov, and P.~Lemonde.
\newblock Observation and cancellation of a perturbing dc {S}tark shift in
  strontium optical lattice clocks.
\newblock {\em Ultrasonics, Ferroelectrics, and Frequency Control, IEEE
  Transactions on}, 59(3):411--415, 2012.

\bibitem{PhysRevLett.110.180802}
K.~Gibble.
\newblock Scattering of cold-atom coherences by hot atoms: Frequency shifts
  from background-gas collisions.
\newblock {\em Phys. Rev. Lett.}, 110:180802, 2013.

\bibitem{guena2012}
J.~Guéna, M.~Abgrall, D.~Rovera, P.~Laurent, B.~Chupin, M.~Lours,
  G.~Santarelli, P.~Rosenbusch, M.E. Tobar, R.~Li, K.~Gibble, A.~Clairon, and
  S.~Bize.
\newblock Progress in atomic fountains at {LNE-SYRTE}.
\newblock {\em Ultrasonics, Ferroelectrics, and Frequency Control, IEEE
  Transactions on}, 59(3):391--409, 2012.

\bibitem{guena2010}
J.~Guéna, P.~Rosenbusch, P.~Laurent, M.~Abgrall, D.~Rovera, G.~Santarelli,
  M.E. Tobar, S.~Bize, and C.~Clairon.
\newblock Demonstration of a dual alkali {R}b/{C}s fountain clock.
\newblock {\em Ultrasonics, Ferroelectrics, and Frequency Control, IEEE
  Transactions on}, 57(3):647--653, 2010.

\bibitem{guena2011}
J.~Gu\'ena, R.~Li, K.~Gibble, S.~Bize, and A.~Clairon.
\newblock Evaluation of {D}oppler shifts to improve the accuracy of primary
  atomic fountain clocks.
\newblock {\em Phys. Rev. Lett.}, 106:130801, 2011.

\bibitem{abgrall2015461}
M.~Abgrall, B.~Chupin, L.~De Sarlo, J.~Guéna, P.~Laurent, Y.~Le Coq, R.~Le
  Targat, J.~Lodewyck, M.~Lours, P.~Rosenbusch, G.~D. Rovera, and S.~Bize.
\newblock Atomic fountains and optical clocks at {SYRTE}: Status and
  perspectives.
\newblock {\em Comptes Rendus Physique}, 16(5):461 -- 470, 2015.

\bibitem{guena2014}
J.~Guéna, M.~Abgrall, A.~Clairon, and S.~Bize.
\newblock Contributing to {TAI} with a secondary representation of the {SI}
  second.
\newblock {\em Metrologia}, 51(1):108, 2014.

\bibitem{Zhang2012}
W.~Zhang, M.~Lours, M.~Fischer, R.~Holzwarth, G.~Santarelli, and Y.~Le~Coq.
\newblock Characterizing a fiber-based frequency comb with electro-optic
  modulator.
\newblock {\em Ultrasonics, Ferroelectrics, and Frequency Control, IEEE
  Transactions on}, 59(3):432--438, 2012.

\bibitem{CIPM}
Recommendation 2 ({CI}-2015).
\newblock {\em Session II of the 104th meeting of the CIPM}, 2015.

\bibitem{baillard2008optical}
X.~Baillard, M.~Fouch{\'e}, R.~Le~Targat, P.~G. Westergaard, A.~Lecallier,
  F.~Chapelet, M.~Abgrall, G.~D. Rovera, P.~Laurent, P.~Rosenbusch, et~al.
\newblock An optical lattice clock with spin-polarized $^{87}\mathrm{Sr}$
  atoms.
\newblock {\em The European Physical Journal D}, 48(1):11--17, 2008.

\bibitem{0026-1394-45-5-008}
G.~K. Campbell, A.~D. Ludlow, S.~Blatt, J.~W. Thomsen, M.~J. Martin, M.~H.~G.
  de~Miranda, T.~Zelevinsky, M.~M. Boyd, J.~Ye, S.~A. Diddams, T.~P. Heavner,
  T.~E. Parker, and S.~R. Jefferts.
\newblock The absolute frequency of the $^{87}${S}r optical clock transition.
\newblock {\em Metrologia}, 45(5):539, 2008.

\bibitem{PhysRevLett.100.140801}
S.~Blatt, A.~D. Ludlow, G.~K. Campbell, J.~W. Thomsen, T.~Zelevinsky, M.~M.
  Boyd, J.~Ye, X.~Baillard, M.~Fouch\'e, R.~Le~Targat, A.~Brusch, P.~Lemonde,
  M.~Takamoto, F.-L. Hong, H.~Katori, and V.~V. Flambaum.
\newblock New limits on coupling of fundamental constants to gravity using
  $^{87}\mathrm{Sr}$ optical lattice clocks.
\newblock {\em Phys. Rev. Lett.}, 100:140801, 2008.

\bibitem{Hong:09}
F.-L. Hong, M.~Musha, M.~Takamoto, H.~Inaba, S.~Yanagimachi, A.~Takamizawa,
  K.~Watabe, T.~Ikegami, M.~Imae, Y.~Fujii, M.~Amemiya, K.~Nakagawa, K.~Ueda,
  and H.~Katori.
\newblock Measuring the frequency of a {S}r optical lattice clock using a 120
  km coherent optical transfer.
\newblock {\em Opt. Lett.}, 34(5):692--694, 2009.

\bibitem{0026-1394-48-5-022}
S.~Falke, H.~Schnatz, J.~S. R.~Vellore Winfred, T.~Middelmann, S.~Vogt,
  S.~Weyers, B.~Lipphardt, G.~Grosche, F.~Riehle, U.~Sterr, and C.~Lisdat.
\newblock The $^{87}${S}r optical frequency standard at {PTB}.
\newblock {\em Metrologia}, 48(5):399, 2011.

\bibitem{1882-0786-5-2-022701}
A.~Yamaguchi, N.~Shiga, S.~Nagano, Y.~Li, H.~Ishijima, H.~Hachisu, M.~Kumagai,
  and T.~Ido.
\newblock Stability transfer between two clock lasers operating at different
  wavelengths for absolute frequency measurement of clock transition in
  $^{87}${S}r.
\newblock {\em Applied Physics Express}, 5(2):022701, 2012.

\bibitem{Matsubara:12}
K.~Matsubara, H.~Hachisu, Y.~Li, S.~Nagano, C.~Locke, A.~Nogami, M.~Kajita,
  K.~Hayasaka, T.~Ido, and M.~Hosokawa.
\newblock Direct comparison of a {C}a$^+$ single-ion clock against a {S}r
  lattice clock to verify the absolute frequency measurement.
\newblock {\em Opt. Express}, 20(20):22034--22041, 2012.

\bibitem{1882-0786-7-1-012401}
D.~Akamatsu, H.~Inaba, K.~Hosaka, M.~Yasuda, A.~Onae, T.~Suzuyama, M.~Amemiya,
  and F.-L. Hong.
\newblock Spectroscopy and frequency measurement of the $^{87}${S}r clock
  transition by laser linewidth transfer using an optical frequency comb.
\newblock {\em Applied Physics Express}, 7(1):012401, 2014.

\bibitem{doi:10.7566/JPSJ.84.115002}
T.~Tanabe, D.~Akamatsu, T.~Kobayashi, A.~Takamizawa, S.~Yanagimachi,
  T.~Ikegami, T.~Suzuyama, H.~Inaba, S.~Okubo, M.~Yasuda, F.-L. Hong, A.~Onae,
  and K.~Hosaka.
\newblock Improved frequency measurement of the $^1{S}_0 - {}^3{P}_0$ clock
  transition in $^{87}${S}r using a {C}s fountain clock as a transfer
  oscillator.
\newblock {\em Journal of the Physical Society of Japan}, 84(11):115002, 2015.

\bibitem{0256-307X-32-9-090601}
L.~Yi-Ge, W.~Qiang, L.~Ye, M.~Fei, L.~Bai-Ke, Z.~Er-Jun, S.~Zhen, F.~Fang,
  L.~Tian-Chu, and F.~Zhan-Jun.
\newblock First evaluation and frequency measurement of the strontium optical
  lattice clock at {NIM}.
\newblock {\em Chinese Physics Letters}, 32(9):090601, 2015.

\end{thebibliography}

\end{document}